\documentclass[a4paper,11pt]{article} %

\usepackage[a4paper,top=1.2cm,bottom=1.8cm,left=1.4cm,right=1.4cm,marginparwidth=1.75cm]{geometry}

\usepackage{lineno}
\usepackage[T1]{fontenc}
\usepackage[utf8x]{inputenc}
\usepackage[UKenglish]{babel}

\usepackage{amsmath}
\usepackage{graphicx}
\usepackage{ulem}
\usepackage[colorinlistoftodos]{todonotes}
\usepackage{url}
\usepackage{xfrac}
\usepackage{cite}
\everypar{\looseness=-1}

\usepackage[colorlinks=true, allcolors=blue]{hyperref}

\title{The "DIS and Related Subjects" Strategy Document:\\ Fundamental Science from Lepton-Hadron Scattering}
\author{Contact persons:
  Allen Caldwell\rlap,$^1$
  Rolf Ent\rlap,$^2$
  Aharon Levy\rlap,$^3$
  Paul Newman$^4$ and
  Fred Olness$^5$
  \\[2ex]   \small \it
${}^1$Max Planck Institute for Physics, Munich, Germany
\\   \small \it
${}^2$Thomas Jefferson National Accelerator Facility, Newport News, VA 23606, USA
\\     \small \it
${}^3${Raymond \& Beverly Sackler School of Physics \& Astronomy, Tel Aviv University, Tel Aviv, Israel}
\\   \small \it
${}^4$School of Physics and Astronomy, University of Birmingham, Birmingham, UK
 \\    \small \it
 ${}^5$Department of Physics, Southern Methodist University, Dallas, TX 75275 USA 
 \vspace{2.0cm}
}

\begin{document}
\maketitle
\begin{abstract}
The diverse community of scientists involved in Deep Inelastic
Scattering includes about 2000 experimental and theoretical physicists
worldwide and envisages projects such as the EIC, LHeC, FCC-eh
and VHEeP as future lepton-hadron scattering facilities.
The proposed facilities will address fundamental questions in strong
interaction / QCD physics, including a first-ever tomographic mapping of the
hadrons' internal structure, a solution to the proton mass and spin problems,
understanding of the high-energy structure of hadronic matter and insight into
connections between gravity and the strong interactions.
They also extend and enhance the CERN programme of searches for new physics
at the energy frontier, through a standalone precision Higgs and top
programme, considerable sensitivity to the direct production of new particles
and the most precise determinations of proton and nuclear structure in the
kinematic range that is relevant to the LHC.
In particular, we highlight the complementary aspects of the different
lepton-hadron projects, and underscore how all are required to provide
a complete characterization of the physics across the full kinematic
reach.
This review of the proposed
facilities and their vast potential for particle physics was inspired
by the discussions in the 2018 ``DIS and Related Subjects” Workshop.
\end{abstract}

\pagenumbering{gobble} 
\newpage
\setcounter{page}{1} 
\pagenumbering{arabic}

\section{Scientific Framework}

The road to new fundamental physics may take an unexpected direction; we should consider them all.
As well as the searches at the energy frontier, a deeper understanding of the particles and forces that are already known could also bring new and exciting physics.
Deep Inelastic Scattering (DIS) provides an excellent framework to make a decisive and transformational 
leap in our understanding of the strong interaction, while also providing discovery potential 
via precision measurements and sensitivity to the direct production of new particles. 
{\bf
The new lepton-hadron facilities described in this document and summarized in Table~\ref{tab:facilities}
below offer dramatically new energy 
and luminosity reach in the DIS context, as well as a full set of polarizations and a vast range of heavy ions 
in addition to proton beams, representing an exciting and
diverse programme for the 
medium and long-term future.
}

The Large Hadron electron Collider (LHeC) and FCC-eh projects~\cite{AbelleiraFernandez:2012cc}
offer $ep$ and $eA$ collisions at TeV-scale  center-of-mass energy and  
luminosities of order 1000 times larger than that of HERA, the first and to date the only lepton-hadron
collider worldwide, providing fascinating probes of QCD and hadron structure as well as a novel
configuration for Higgs and Beyond the Standard Model (BSM) physics. With a similar three orders of magnitude
increase in luminosity and polarized hadron beams, the  Electron Ion Collider (EIC)~\cite{Accardi:2012qut}
offers unique insights into nucleon and nuclear spin and tomography, emphasizing gluon degrees of freedom. 
Recently, the possibility of a very high energy, modest luminosity, electron-proton/ion collider (VHEeP) \cite{Caldwell:2016cmw}
has been raised based on plasma wakefield acceleration, which could extend the center-of-mass energy to
close to $10 \ {\rm TeV}$ into a region where new strong interaction dynamics must exist and connections
with high-energy cosmic rays and gravitational physics may be strongest.
An early version using the SPS to accelerate electrons is also discussed (PEPIC).
Detailed science cases and synergies can be found in the separate documents submitted related to the LHeC, EIC and VHEeP.

\begin{table*}[h!]
\begin{footnotesize}
\begin{center}
\caption{Overview of proposed electron-hadron colliders.}
\begin{tabular} {|l|l|l|l|l|l|} \hline
Facility& Years& $E_{cm}$ & Luminosity& Ions & Polarization \\ 
 & & (GeV) & ($10^{33}cm^{-2}s^{-1}$) & & \\ \hline
EIC in US & $>2028$ & 20 - 100 $\to$ 140 & 2 - 30 & p $\to$ U & e, p, d, $^3$He, Li\\
EIC in China & $>2028$ & 16 - 34 & 1 $\to$ 100 & p $\to$ Pb& e, p, light nuclei\\ 
LHeC (HE-LHeC) & $>2030$ & 200 - 1300 (1800) & 10 & depends on LHC & e possible \\
PEPIC & $>2025$ & 530 $\to$ 1400 & $< 10^{-3}$ & depends on LHC & e possible \\
VHEeP & $>2030 $ & 1000 - 9000 & $10^{-5} - 10^{-4}$ & depends on LHC  & e possible \\
FCC-eh & $>2044 $& 3500 & 15 & depends on FCC-hh & e possible \\ \hline
\end{tabular}
\end{center}
\end{footnotesize}
\label{tab:facilities}
\end{table*}

The physics highlights available from combining the capabilities of these facilities are:
\begin{itemize}
\item Understanding how the features of protons, neutrons and other hadrons such as their mass, spin and 3D structure emerges from their quark and gluon constituents;
\item Understanding the fundamental state and modes of interaction of these constituents at high-energy  and exploring the relation of this physics to gravitational physics;
\item Expansion of the CERN hadron-hadron collider via high precision Higgs measurements;
\item Maximize the Beyond the Standard Model physics search potential at high energies by exploiting the unique capabilities of an $ep$ collider;
\item Precision descriptions of protons and ions over a vast kinematic range allowing for discoveries at the LHC and future colliders.
\end{itemize}
Among the proposed projects discussed here, the EIC is perhaps furthest along to approval,
having recently received strong support from the National Academy of Sciences of the USA. Detailed studies
and conceptual and technical designs have been prepared for the LHeC, and technology developments of energy
recovery linacs, a central element in the LHeC/FCC-eh projects, are underway.
VHEeP and PEPIC are in early conceptual stages, but rapid progress can occur given further successful demonstrations of the plasma acceleration technology.
The physics capabilities and objectives of these facilities are further explored in sections~\ref{qcdsec}, 
\ref{precisionsec} and~\ref{bsmsec}. Their technical realization and readiness are 
discussed in section~\ref{methodology}, and we summarize in section~\ref{summary}.
For consistency, we generally discuss the facilities in order of increasing energy; this does not indicate a ranking or scientific judgment.

\section{A New Understanding of QCD}
\label{qcdsec}

Of the fundamental forces within the current Standard Model of particle physics, 
QCD is the most complex and enigmatic. The strong coupling and the non-Abelian 
self-interacting properties of QCD manifest themselves in a theory which displays both 
asymptotic freedom (at short distance scales) and confinement of the quarks and gluons 
(at large distance scales). These characteristics make QCD 
impossible to formulate perturbatively at the low energies characteristic of hadronic bound states.
The confinement property implies that 
under almost all circumstances, gluons manifest collectively, the consequences of which are profound, 
but very poorly understood. 
It is deeply important theoretically and challenging experimentally to 
understand how gluons bind the nucleon together and generate the strong interaction field 
energy that provides hadronic mass, and 
to unravel the complex dynamics of quark and gluon polarizations and orbital angular 
momenta that produce the exact spin--$\,\sfrac{1}{2}$
nature of the nucleon.  
In his 2013 DIS Workshop summary, Chris Quigg reminded us that 
``our understanding of QCD is not complete'' as evidenced 
by the unresolved strong CP problem, and it is ``highly likely that
we may find new phenomena  {\it  within} QCD.''
{\bf
A deeper comprehension of QCD lies at the heart of 
our exploration of the Standard Model and beyond;
}
this endeavor is synergistic with a broad spectrum of research as diverse as 
dark matter searches and electroweak symmetry breaking~\cite{Quigg:2013lya}.

\subsection{Emergence of Basic Hadronic Properties from Collective effects}
\label{sec:emergence}

The facts that the proton is massive, apparently absolutely stable and has spin--$\,\sfrac{1}{2}$ are 
fundamental to our understanding of the universe.
Yet these are not basic 
principles of the Standard Model, but are emergent phenomena in which simple fundamental
rules of nature are produced from multi-body problems via 
non-perturbative mechanisms. These
are targets that once seemed intractable from both the experimental and the theoretical 
standpoints.
{\bf
The future DIS facilities discussed here would together yield a precise mapping of partons and the correlations between them over a vast kinematic range, in particular with a new level of sensitivity to the role of gluons, which is likely to be key.
}
``Understanding the Glue'' is the central vision of the EIC project, to be tackled via new
approaches to observing and characterizing the gluons in hadrons at intermediate and large Bjorken-$x$,
including fully determining the polarized gluon density and understanding how hadron masses and spin emerge from
the underlying quark and gluon constituents, and their interactions.
LHeC covers a much larger range in $x$ and much higher scale $Q^2$, where a fully perturbative approach is
readily applicable, and has a chance to characterize and prove saturation to exist, using beams of
protons and (most likely) lead ions. The very low Bjorken-$x$ regime, below 10$^{-6}$ will be accessible
by FCC-eh, which extends to 10~TeV$^2$ in $Q^2$, and by VHEeP.

The strongly interacting nature of QCD gives rise to unique phenomena under extreme 
conditions such as the formation of a quark-gluon plasma. 
In the corresponding experimental
programme of heavy ion collisions, 
it is vital to establish a baseline for behavior in the absence of
plasma formation. However,
there are novel collective effects such as 
unexpected ridge structures in proton-lead \cite{Abelev:2012ola,Aad:2012gla,Aaij:2015qcq,CMS:2012qk}
and even proton-proton \cite{Khachatryan:2010gv} collisions, which have only begun to be explored.
Recently, a surprising universal pattern in the particle ratio independent of the collision type was demonstrated.~\cite{Acharya:2018orn}
{\bf DIS of leptons on protons and heavy ions across all of the proposed future facilities will yield 
a new level of understanding of the initial state of heavy ion collisions}~\cite{Kovarik:2015cma,Eskola:2016oht}
and a baseline to compare with new observations in their final states.

\subsection{Beyond the Collinear Parton Model}

The characterization of hadrons in the framework of the parton model has been remarkably 
successful. Most of our current knowledge of hadronic structure and parton dynamics is in 
the form of their number- or momentum-density as a function of Bjorken-$x$ and scale $Q^2$. 
Here, the picture is of a collection of fast-moving partons, with transverse degrees of 
freedom and inter-parton correlations ignored. Whilst this longitudinal 
information is vitally important in many contexts, notably for understanding the very 
high energy collisions at the LHC (section~\ref{precisionsec}), 
it leaves a number of fundamental questions about 
structure unanswered. As we push to higher precision (NNLO and beyond) and 
extreme kinematic regions, the limitations of this framework are revealed 
in the form of higher-twist contributions, collective phenomena, and non-factorisable contributions. 
More fundamentally, 
the description of hadronic structure in terms of only the longitudinal momentum fraction 
is a drastic over-simplification; this picture neglects the rich 3-dimensional tomographic
structure characterizing the transverse and spatial distributions of the partons within hadrons.
\goodbreak 
By exploiting its potential to polarize protons and light ions both longitudinally and transversely,
the EIC will provide the opportunity to 
{\bf
go far beyond the current one-dimensional picture, 
enabling nucleon and nuclear ‘femtography’ by correlating the information of the quark and gluon 
longitudinal momentum component with its transverse degrees of freedom,
}
through measurements of transverse momentum dependent parton densities and 
(through exclusive processes such as deeply-virtual Compton scattering) generalized parton densities.
The energies of the EIC ($\sqrt{s}$ = 20 - 100 GeV, upgradable to 140 GeV), are well matched to 
tackle this problem at the intermediate-to-large momentum fractions that are driven by
the underlying valence structure of nucleons and nuclei
and the simplest quantum fluctuations.
\begin{figure}[!ht]
\begin{center}
\includegraphics[width=0.99\textwidth]{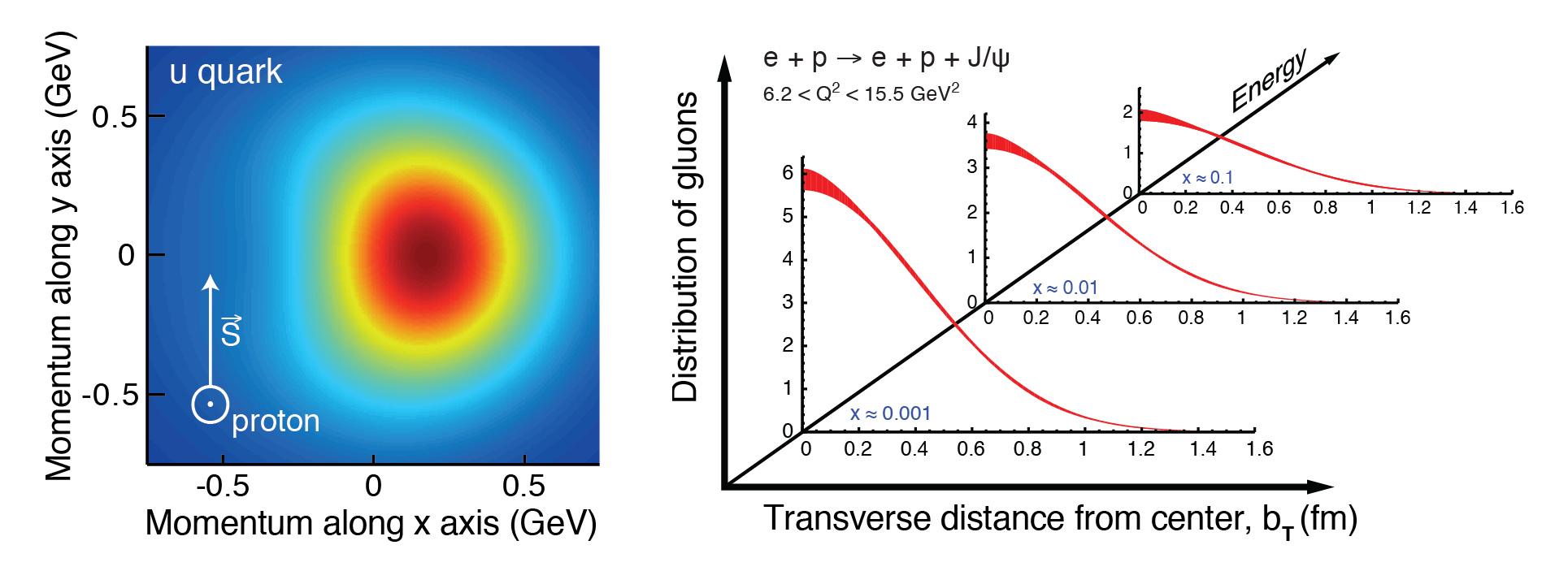}
\caption{
(Left) A simulation based on projected EIC data of the transverse motion preferences of an up sea quark within a proton moving out of the page, with its spin pointing upward. 
The color code indicates the probability of finding the up quarks. 
\quad 
(Right) Projected measurement precision of the distribution of gluons in transverse space inside a proton, obtained from exclusive J/$\psi$ production at the EIC. Projections are shown for three different bins in the fraction of the proton's momentum carried by the gluons~\cite{Accardi:2012qut}.}
\label{fig:EICtrans}
\end{center}
\end{figure}

Figure~\ref{fig:EICtrans} illustrates the new sensitivity to 3-dimensional hadronic 
structure offered by the EIC \cite{EIC:LRP}. Figure~\ref{fig:EICtrans} (left) shows a simulation
of how transverse anisotropies might be revealed in semi-inclusive polarized DIS processes
in which the scattered electron and struck parton directions are both reconstructed, 
with corresponding sensitivity to the ingredients of proton spin. 
Figure~\ref{fig:EICtrans} (right) shows how the transverse spatial structure of the gluon density
might also be measured through exclusive processes (in this case $ep \to e J/\psi p$).

The LHeC has the 
advantage of vastly enhanced kinematic
coverage at small $x$ and high $Q^2$.
The low $x$ coverage in particular
would provide crucial information to deepen our understanding
of the spatial and momentum distribution of partons in unpolarized protons and nuclei~\cite{AbelleiraFernandez:2012cc,Armesto:2014sma}.

The high luminosity provided by the EIC and LHeC will further allow  to understand and quantify the role of multi-parton interactions in probing both structure and reaction mechanism. In QCD and in DIS, an energetic parton 
can interact within a hadron multiple times before exiting 
to hadronize.
This type of multi-parton interaction provides unique opportunities to probe multi-parton correlations. 
The hard scattering of a lepton off a hadron or a nucleus 
provides the cleanest, best controlled probe to access 
multi-parton correlations, 
and to establish their role in the interpretation of the transverse momentum dependent parton distribution functions.
With the high luminosities promised at the new colliders, studying parton entanglement will also 
become possible~\cite{Kharzeev:2017qzs}.

\subsection{Novel QCD Dynamics at Small Bjorken-x}
\label{sec:saturation}

As the Bjorken-$x$ becomes very small, a regime is entered in which QCD radiation is the dominant feature and the fundamental parton splitting processes that lie at
the heart of the strong interaction are therefore probed.
Conventionally the DGLAP equations (Dokshitzer-Gribov-Lipatov-Altarelli-Parisi)~\cite{Gribov:1972ri, Gribov:1972rt, Lipatov:1974qm, Dokshitzer:1977sg,Altarelli:1977zs} prescribe the
 $Q^2$~evolution that has been the basis of global fits to extract parton densities. At smaller $x$,
the BFKL equation
(Balitsky-Fadin-Kuraev-Lipatov)~\cite{Balitsky:1978ic,Kuraev:1976ge} becomes valid.
Including $\log(1/x)$ (BFKL) resummation~\cite{Ball:2017otu} has recently been
shown to improve the description of HERA data at the lowest $x$ values.
As discovered at HERA, the low $x$ region of proton structure is 
characterized by a rapid increase in the density of partons for larger scales $Q^2$. It is a natural, but as 
yet not fully resolved, question as to whether this rise continues to indefinitely small $x$, 
or whether, like the $WW$ scattering cross section, there must be a 
new mechanism at high energy to tame the growth 
and satisfy unitarity constraints. A proposed solution in perturbative partonic language is
the recombination of pairs of gluons ($gg \to g$) when they are sufficiently densely 
packed to `saturate’. 

Due to the enhancement in target density with increasing atomic number ($\sim \! A^{1/3}$), 
nuclei provide small~$x$ observables which are complementary to those available in DIS from 
protons and hence a naturally increased sensitivity to saturation phenomena.
Indeed, changing $A$, {\sl i.e.} measurements with different ions, is a new axis along which non-linear
effects and the approach to the saturation phenomenon are expected to become important.
The wide range of ions expected to be available at the EIC will offer unique opportunities for these studies.
The polarized $ep$ program at an EIC similarly can look for double-logarithmic effects possibly generating
the small-$x$ behavior of polarized structure functions, a strong and clear signature of BFKL-like effects
~\cite{Bartels:1996wc,Kovchegov:2016weo}, not expected or found anywhere else.

\begin{figure}[!ht]
\begin{center}
\includegraphics[width=0.90\textwidth]{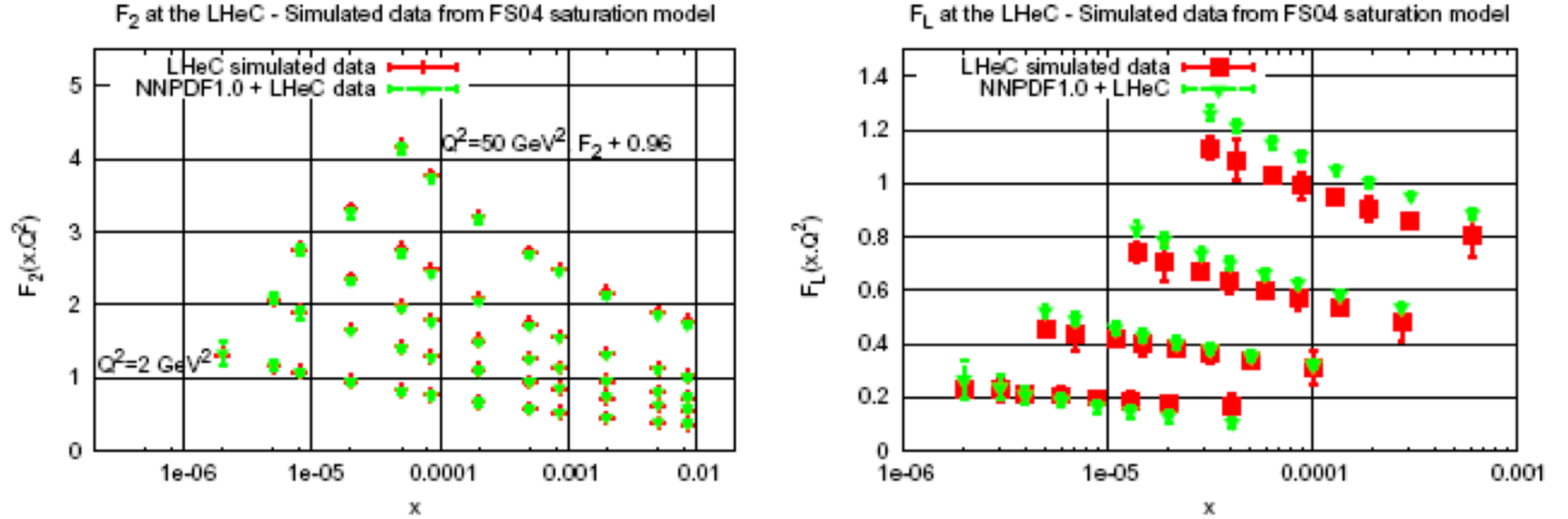}
\vspace*{-0.4cm}
\caption{Simulation of LHeC $F_2$ and $F_L$ proton structure function measurements (red),
generated according to the Forshaw \& Shaw 2004 dipole model \cite{Forshaw:2004vv}, in which saturation effects are included based on fits to low $Q^2$ HERA data. The green points show the best fit obtained when these simulated data 
are added into the NNPDF 1.0 global fitting framework, which is
based on NLO QCD with no low $x$ effects included. Figure 
from \cite{AbelleiraFernandez:2012cc}.}
\label{fig:f2fl}
\end{center}
\vspace*{-0.4cm}
\end{figure}

With its higher center of mass energy, the LHeC (and PEPIC)
accesses small $x$ physics in the region where 
partons are expected to exhibit saturated behavior.
Combining information from 
multiple observables in inclusive and diffractive $ep$ and $eA$ 
scattering is a powerful means
not only of establishing the presence of 
saturation effects, but also differentiating between models.
An example based on simulated inclusive LHeC $ep$ data is shown 
in figure~\ref{fig:f2fl}. By scanning the electron beam (and hence the
 center-of-mass) energy, the proton structure 
functions $F_2(x,Q^2)$ and $F_L(x,Q^2)$ are both extracted over a wide range 
of 
$Q^2$ values down to $x$ values below $10^{-5}$. 
Clear evidence for saturation could be obtained
in the form of a
tension between the 
direct gluon sensitivity of $F_L$ and that through the 
logarithmic $Q^2$ dependence of $F_2$.
Whatever the relevant new dynamics, 
the high precision achievable with the LHeC will provide crucial input into understanding small-$x$ physics.

\begin{figure}[!ht]
\begin{center}
\includegraphics[width=0.6\textwidth]{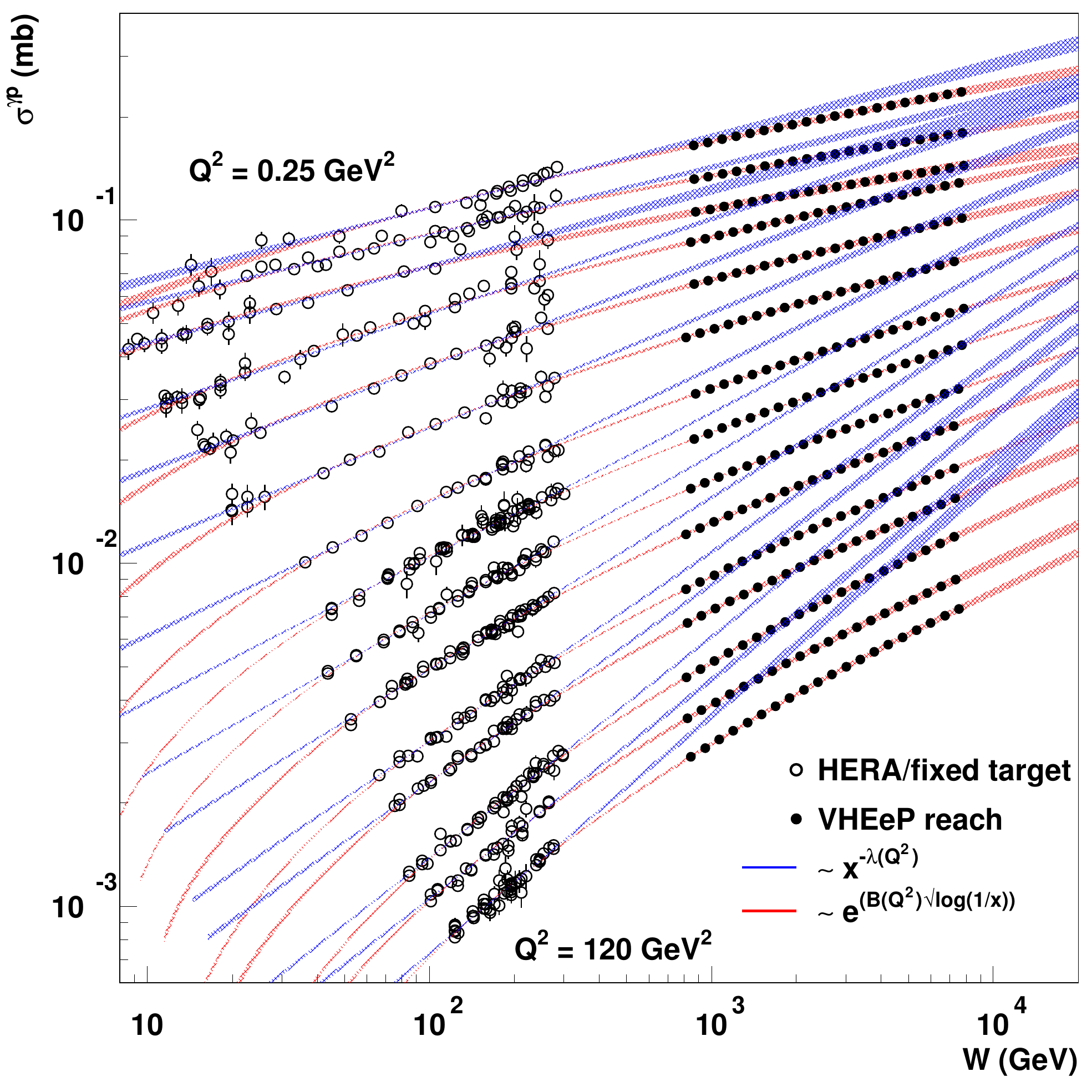}
\caption{Measurements (open points) of $\sigma_{\gamma p}$ versus $W$ for $0.25 < Q^2 < 120$ GeV$^2$ from HERA and fixed-target experiments. The blue and red lines show different fits to the data. The kinematic reach of VHEeP is shown as projected data points (closed points). The uncertainties are assumed to be of order 1\%, given the increased cross section expected and similar systematics to those at HERA and are not visible as error bars on this plot. LHeC and PEPIC would cover the kinematic range between HERA and VHEeP.}
\label{fig:sig-W}
\end{center}
\end{figure}

A further step towards still lower $x$ can be taken with FCC-eh and VHEeP, where $W=\sqrt{Q^2/x}$ values of $2\cdot 10^3 - 10^4$~GeV would be reachable, corresponding to $x$ of $10^{-7}-10^{-8}$ for $Q^2 = 1 \ {\rm GeV^2}$.  With its very large increase in kinematic range, VHEeP may be the ultimate facility for this physics programme, where cross sections are typically large and luminosity is correspondingly not a major issue. 
Figure~\ref{fig:sig-W} shows the results of extrapolation of fits~\cite{Caldwell:2016cmw} to the 
center-of-mass energy dependence of the photon–proton cross section for
a range of scales $Q^2$ and two
assumptions on the small $x$ behavior of the parton distributions.
In one instance (blue curves), the cross section is assumed to follow a simple power-law behavior at 
small values of $x$
while in the second instance (red curves) 
a form inspired by double asymptotic ${x,Q^2}$ scaling~\cite{Ball:1994du} was used.
Whilst current data are compatible with both of these forms, they diverge dramatically in the TeV energy
range, illustrating just how 
little is known about the high energy behavior of hadronic cross sections. 

{\bf The measurements 
that are possible at future
facilities across a range of energies and virtualities 
will clearly yield exciting and unique information on the 
understanding of the strong interaction 
and reveal the fundamental underlying physics at the heart of high energy hadronic cross sections.}

\subsection{QCD and Gravity}

The connection between gravity and the Standard Model forces remains mysterious. There are intriguing theoretical links between gravity and QCD: the well-known AdS/CFT conjecture links gravity-like theories to QCD-like strong coupling theories~\cite{Maldacena:1997re}.
This conjecture has spurred tremendous activity and theoretical tools based on this relation have led
to successes such as the calculation of the ratio of the shear viscosity and volume density of entropy
in a quark-gluon plasma~\cite{Kovtun:2004de}.
In the AdS/CFT-based theoretical studies, gravitational interactions are dual to QCD with large coupling constant; i.e, at large distances, where color confinement effects become important, and this connection
is vigorously investigated by several groups of string theorists with
supersymmetry playing a central role \cite{Bern:2008qj,Bern:2010ue,Ammon:2015wua}. Recently, it has been shown that one-graviton
amplitudes can be expressed as linear combinations of pure gluon-like amplitudes \cite{Stieberger:2016lng}.
At this time, this appears to be a mathematical result, but physical consequences could emerge.

The study of DIS at extremely high energies could deliver the relevant data to
bring about a revolution in our understanding of the relationship between the forces.  In the BFKL approach to small-$x$ physics, the high energy behavior of the DIS cross sections is governed by the Pomeron. The Pomeron, a vacuum-like state of QCD, has properties similar to the graviton in gravitational physics; it shows a non-Wilsonian behavior in that short distances are not necessarily connected to high energy scales. E.g., BFKL studies of HERA data~\cite{Kowalski:2017umu} indicate sensitivity to new physics such as supersymmetry, although the scale of supersymmetry may be many, many orders of magnitude higher than the scale of the Pomeron. This is similar to gravitational interactions where the fundamental objects, high mass black holes, grow in size with energy. 

This similarity in the nature of the forces and the fundamental constituents between QCD and gravity and the mathematical connection apparent in the AdS/CFT approach necessitate further investigations.  {\bf Probing the structure of hadronic matter at the highest energy scales possible could yield the fundamental breakthrough in understanding this deep physics.}

\section{Precision Measurements}
\label{precisionsec}

The high luminosities promised by the LHeC and EIC, and the complementarity of the energy reach and polarization, will allow
a new level of precision in measurements of 
proton and nuclear structure. Precise measurements of proton structure will be needed to 
extract the full physics potential from the LHC, while nuclear structure functions
have important implications for the interpretation of heavy ion collisions at the LHC 
and the Brookhaven Relativistic Heavy Ion Collider (RHIC), disentangling initial state 
effects from phenomena occurring during and after the 
collision, as required for a full understanding of the different stages of quark-gluon plasma formation.

\subsection{Proton Structure: enabling LHC discoveries}
\label{lp:precise}

\begin{figure}[!ht]
\begin{center}
\includegraphics[width=0.46\textwidth]{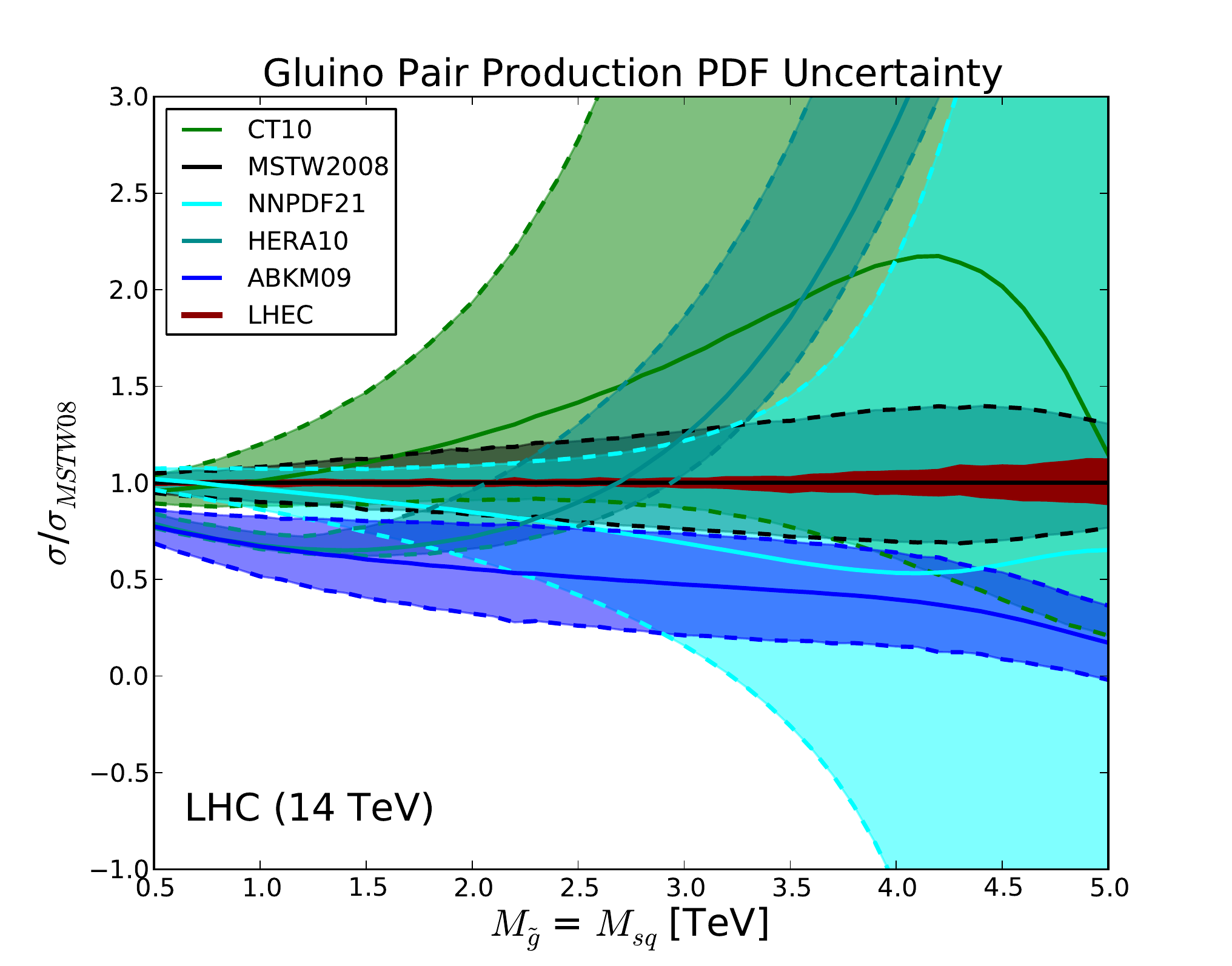}
\includegraphics[width=0.51\textwidth]{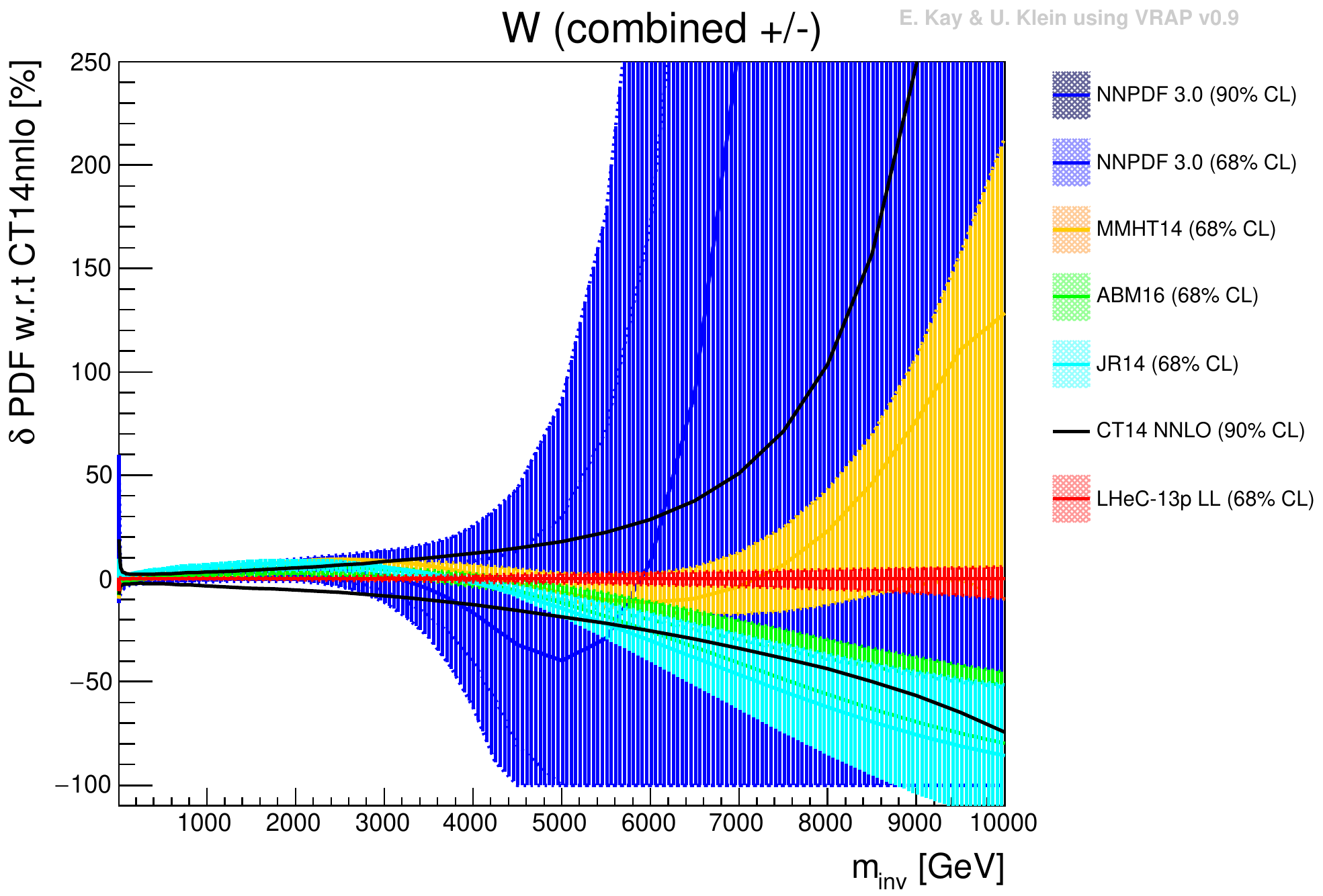}
\vspace*{-0.3cm}
\caption{Uncertainty bands in predicting cross sections for LHC BSM particle
production assuming a fixed non-zero coupling strength for a representative range of PDF sets. (left) gluino pair production via $gg \rightarrow \tilde{g} \tilde{g}$, as a function of the gluino mass \cite{lhec:lhc}. (right) $W^\prime$ boson production via $q \bar{q} \rightarrow W^\prime$ as a function of the $W^\prime$ mass \cite{uta:ellis}.}
\label{fig:bsm}
\end{center}
\end{figure}

As has been exquisitely shown at HERA \cite{Abramowicz:2015mha},
lepton-proton scattering provides uniquely precise and detailed information on the structure of the 
target proton, well matched to the rapidity plateau at the LHC. However, 
as the LHC programme has unfolded and integrated luminosities have become larger, limits on
the masses of new particles have been pushed into the ${\rm TeV}$ range and
many discovery channels are increasingly limited by the precision of the Standard Model 
predictions. These theoretical uncertainties are
driven to a large extent by lack of knowledge
of the proton parton distribution functions at large Bjorken-$x$.
One example is gluino pair production via $gg \rightarrow \tilde{g} \tilde{g}$, where
knowledge of the high $x$ gluon is the limiting factor (figure~\ref{fig:bsm},~left). Similarly, the ultimate limit on high mass
$W$ boson recurrences comes from knowledge of the proton quark 
densities (figure~\ref{fig:bsm},~right). Even for the LHC Higgs
programme, where the incoming partons have $x$ values in the well measured region from HERA
$\sim\! 10^{-2}$, the uncertainties in the gluon distribution are a limiting factor
in predicting cross sections.  

The LHeC and EIC both aim to operate concurrently with the LHC in the 2030's. Both have the 
potential to make substantial further improvements in our knowledge on the parton distribution
functions of the proton in the kinematic range which is most relevant to the LHC $pp$ programme.
Fully clarifying the behavior of parton densities at large $x$ requires proton targets to
avoid the complications introduced by nuclear effects in fixed target data. To avoid the 
influence of higher twist contributions, it also requires relatively large $Q^2$, which is
in any case kinematically associated with high $x$ at large $\sqrt{s}$.
The large lever arm in $x$ and $Q^2$ at the LHeC would be particularly important here.
The high luminosity at LHeC also leads to the copious production of charm, beauty and even top quarks.
Together with the 
complementary sensitivity offered by the exchange of the weak $W$ and $Z$ bosons at high $Q^2$, 
a full flavor decomposition of the quark distributions could be made, replacing the 
assumptions currently imposed in extractions of the parton densities. 

\begin{figure}[!ht]
\begin{center}
 \includegraphics[width=0.40\textwidth,height=0.38\textwidth]{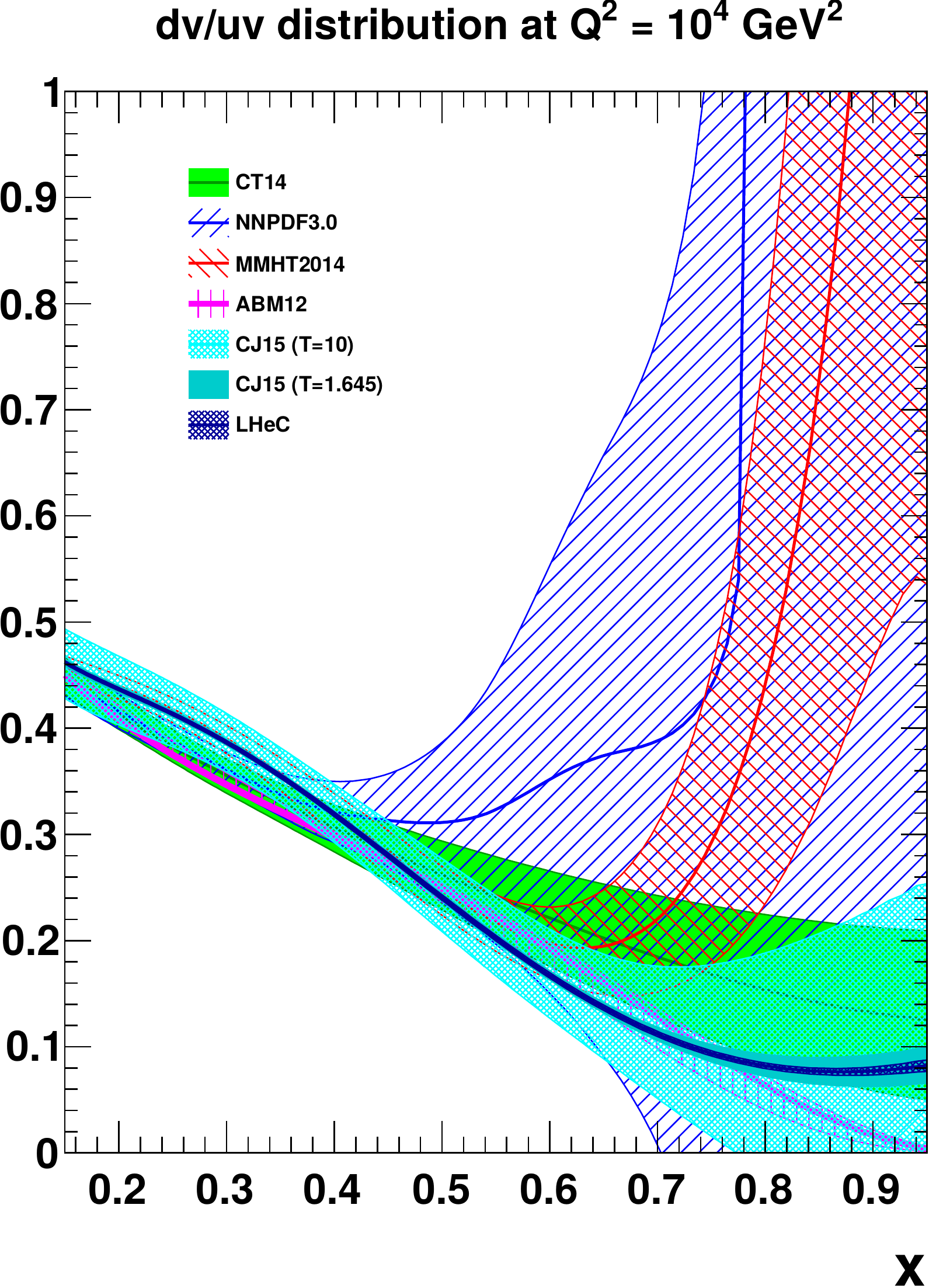}
 \hfil 
\includegraphics[width=0.53\textwidth]{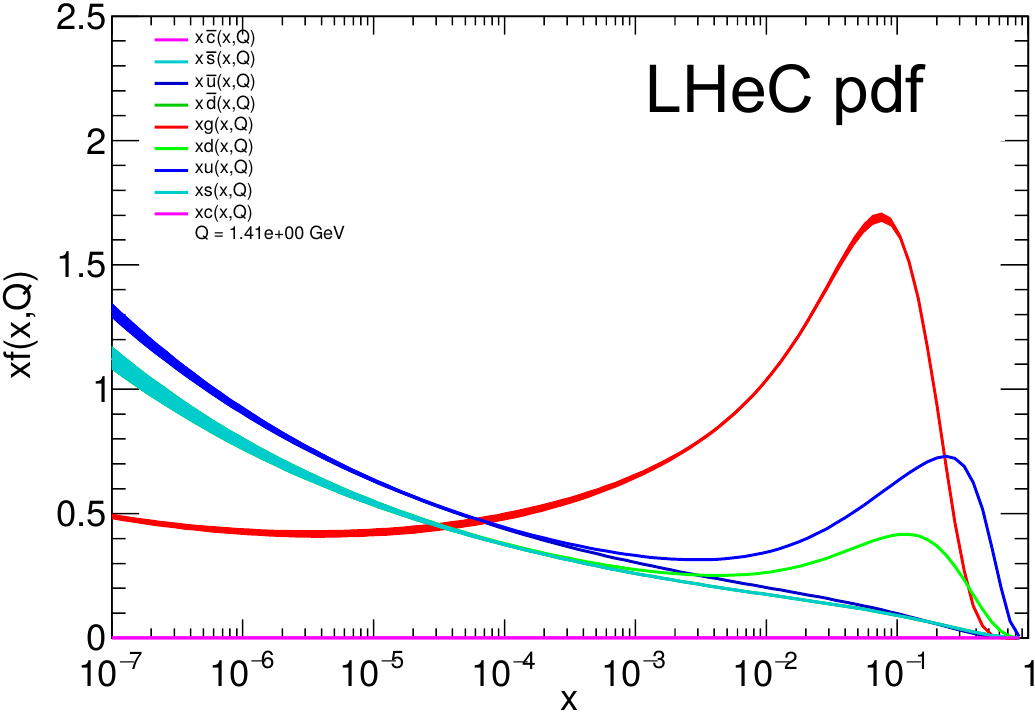}
\vspace*{-0.4cm}
\caption{
a)  Left: Simulated improvement in precision of the ratio of down to up valence quark densities in the large $x$ region from LHeC data at a scale $Q^2$ of 10$^4$ GeV$^2$.
b) Right: Simulation of PDFs and their uncertainty bands with LHeC pseudodata~\cite{dis2018:gwenlan}.
} 
\label{fig:pdfs}
\end{center}
\vspace*{-0.4cm}
\end{figure}

{\bf
Studies in the LHeC context can revolutionize sensitivity to all partons across a very wide range of Bjorken-$x$.
}
As one example, figure~\ref{fig:pdfs}a illustrates the power of the LHeC to resolve the long-standing question
of how the ratio of down to up quarks behaves as $x \rightarrow 1$
in a region of sufficiently high $Q^2$ such that theoretical interpretation is very clean.
The projections make use of the possibility to measure both the neutral- and charged-current
electron-proton scattering process at the high $Q^2$ scales of 10$^4$-10$^5$ GeV$^2$.

The EIC will transform our knowledge
of PDFs at large $x$. One important aspect at the EIC is the possibility of DIS from deuterons
with tagging of spectator protons and neutrons making use of the strong versatility of the EIC beams
and the excellent tagging detection capabilities. The EIC would measure the ratio of down to up quarks
with similar precision as the LHeC in the region close to unity, at $Q^2$ scales of a few-10 to a few-1000 GeV$^2$.
Together, LHeC and EIC will provide a superb constraint of our understanding of evolution at large $x$,
with tremendous lever arm in $Q^2$.

The LHeC also offers exquisite precision on parton densities in the low \hbox{Bjorken-$x$} limit, extending to $\sim\! 10^{-6}$ at perturbative $Q^2$ values  as illustrated in Fig.~\ref{fig:pdfs}b. 
In addition to the 
sensitivity to novel low-$x$ QCD dynamics discussed in section~\ref{sec:saturation}, 
{\bf
this also has direct relevance in
predicting signal and background rates for low mass resonances at the LHC.
In parallel, the strong coupling can be measured to an experimental precision of 0.1\% and the weak mixing angle to~0.2\%.
}

\subsection{Partonic Nuclear Structure}

\begin{figure}[!ht]
\begin{center}
\includegraphics[width=0.99\textwidth]{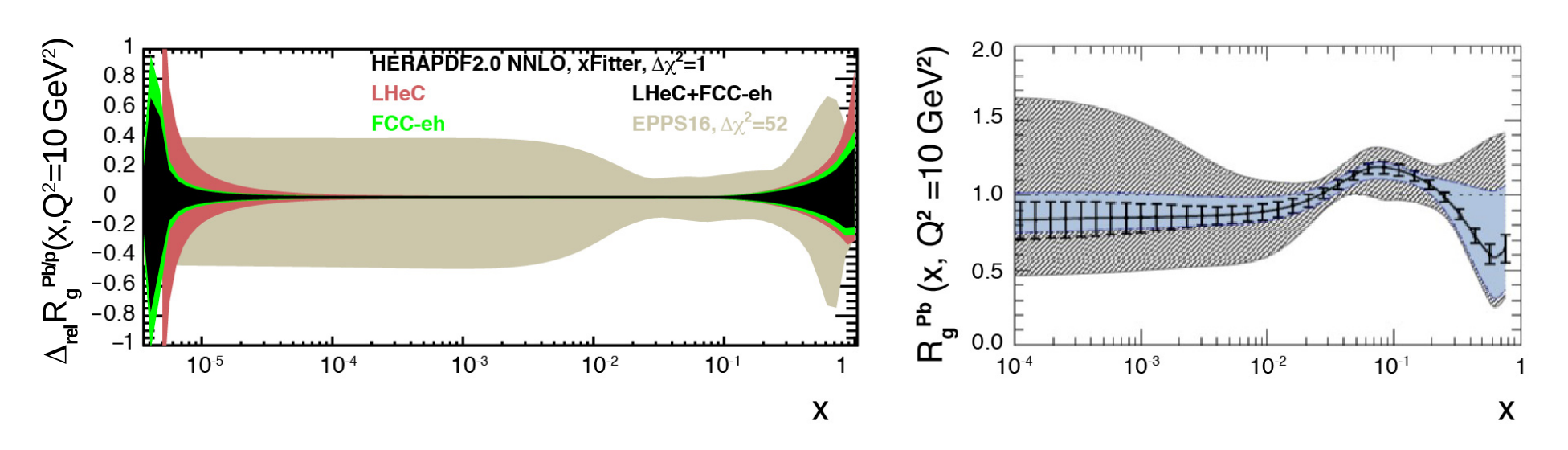}
\vspace{-0.1cm}
\caption{Left: Simulated improvement in the 
precision in the {\sl ratio of} the per-nucleon gluon distributions
between lead-208 and the proton, at $Q^2$ = 10 GeV$^2$, as determined from inclusive structure function fits
of projected LHeC, FCC-eh, and combined data~\cite{Armesto:2018pc}.
The LHeC (FCC-eh) would constrain down to $x$ = 2 $\times$ 10$^{-5}$ (4 $\times$ 10$^{-6}$).
Right: Simulated improvement in the precision of the nuclear modification of the gluon distribution in
lead-208 at $Q^2$ = 10 GeV$^2$ from EIC data. The hatched bands correspond to the baseline fit,
the blue bands are the results of the inclusive structure function fits over a range of $Q^2$,
and the black error bands denote the full analysis with the inclusive and open charm data
~\cite{Aschenauer:2017oxs}. The EIC would constrain down to $x$ = 10$^{-3}$.}
\label{fig:npdfs}
\end{center}
\end{figure}

Electron-ion collisions 
allow to probe the 
nucleon-nucleon interaction in the 
nuclear medium at the highest resolution scales. Up to now, such experiments 
have been limited to low energy fixed target facilities. The nuclear gluon and sea 
quark distributions are thus essentially unknown for $x \leq 10^{-2}$. Results 
from these fixed target experiments have clearly established that the binding of nucleons 
into a nucleus influences the distributions of quarks and gluons. Recent 
experiments at Jefferson Lab have shown that this pattern depends on the nuclear 
structure. An EIC can determine if there is a polarization or quark flavor 
dependence of the pattern of nuclear modifications and map the 
influence of the nuclear medium on the gluon distributions. An LHeC in $ePb$ mode can map 
the nuclear PDFs into an unprecedented low $x$ region, reaching four orders of magnitude beyond
that probed to date,
while the wide range of nuclear beams available at the EIC will offer the ability to study $A$-dependence.

Electron-proton and electron-ion collisions will clarify 
basic questions in %
heavy ion physics as discussed in section~\ref{sec:emergence}.
A detailed picture of
the initial wave function of protons and ions can be obtained, which will settle how small a nuclear system
can become and still show collective features. As one example illustrating the science reach of envisioned
electron-hadron facilities,
figure~\ref{fig:npdfs}(left) shows the tremendous precision and reach in Bjorken-$x$ provided by the LHeC, the
FCC-eh, and the two combined, to constrain the difference 
between the gluon distributions per nucleon in 
lead-208 nuclei and in protons, as determined from the logarithmic dependence of nuclear structure functions
over a large $Q^2$ lever arm. Figure~\ref{fig:npdfs}(right) shows the improvement of the nuclear modification of
the gluon distribution in lead-208 as derived from projected EIC data. Here, the gain in precision at large
Bjorken-$x$ is dominated by the charm-tagged pseudodata.
The EIC will cover the region relevant to RHIC and reach the region of current LHC
heavy-ion experiments.
{\bf 
The LHeC and FCC-eh with their much larger kinematic reach will provide access to  
the complete region of the initial wave function crucial to both LHC and future hadron-hadron and $AA$ colliders,
}
as well as providing
a sizable lever arm in $Q^2$ at small Bjorken-$x$ to
study parton saturation (section~\ref{sec:saturation}).

\label{bsmsec}

With sufficiently large luminosity, a high energy $ep$ collider offers the possibility of copious
production of the Standard Model Higgs boson. With easily identifiable production modes, 
decay modes that have manageable backgrounds and a lack of complications from pile-up, the LHeC 
and FCC-eh are capable of
precision measurements of Higgs couplings, complementing the HL-LHC and future $e^+ e^-$ possibilities. 

Whilst much of the sensitivity of lepton-hadron machines to BSM physics arises indirectly from 
the constraints that they place on Standard Model predictions through 
PDFs (section~\ref{lp:precise}), there are also scenarios for new physics where an initial state
lepton is an advantage, particularly where new particles have couplings to electron plus quark. 

\vspace{-0.15cm}
\subsection{Precision Higgs Measurements}
\label{sec:higgs}
\vspace{-0.15cm}

With many models for new physics predicting deviations of Higgs couplings from Standard Model values 
only at the percent level, the Yukawa couplings to fermions
barely established and not at all in the first and second generations, and the
possibility of 
higher mass recurrences, precision studies of the scalar sector 
will clearly be a pivotal theme in the
future development of particle physics. The Higgs programme 
is therefore a central theme
in the motivation for the high luminosity, high energy LHeC and FCC-eh programmes. 

\begin{figure}[!ht]
\begin{center}
\includegraphics[width=0.95\textwidth]{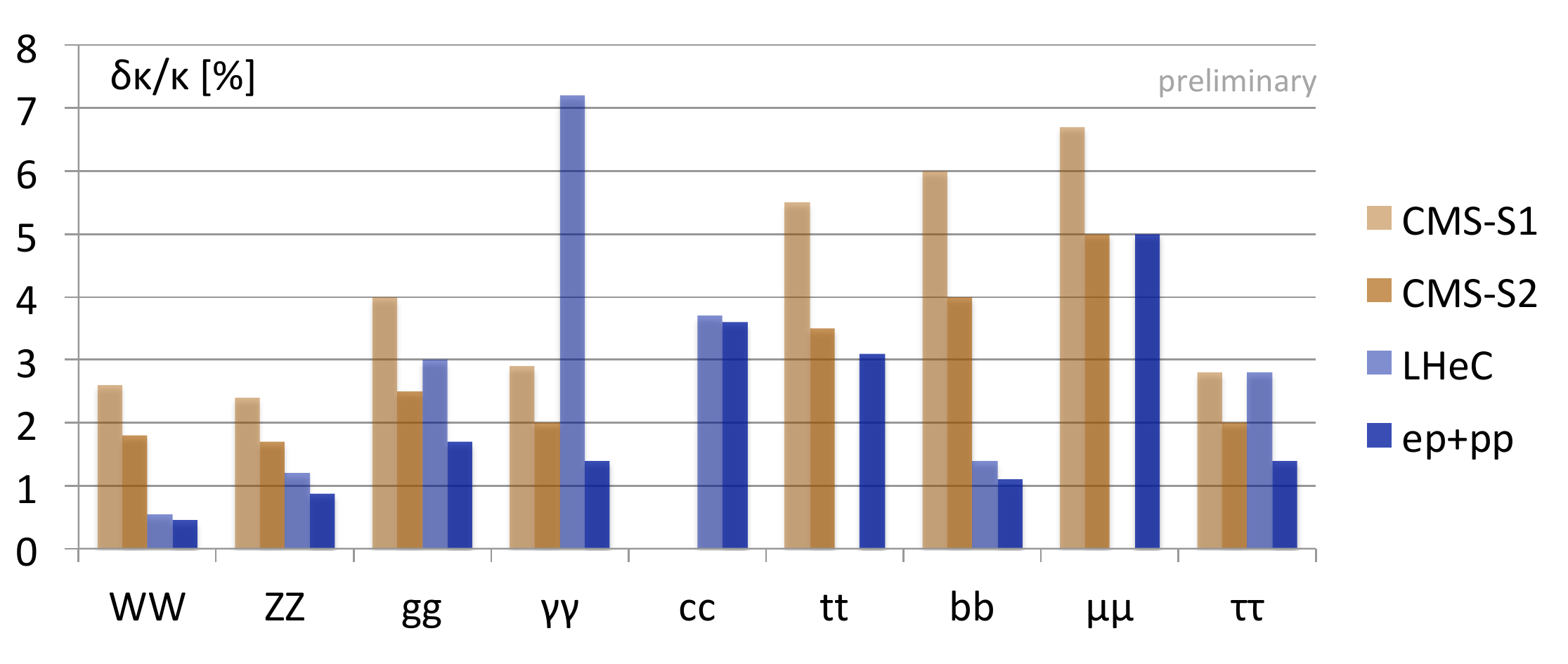}
\caption{ 
Simulated uncertainties on 
Higgs couplings in the kappa formalism.
Determinations are compared
from $3 \ {\rm ab^{-1}}$ in $pp$ at the 
HL-LHC at CMS with conservative (`CMS-S1') and more challenging (`CMS-S2') assumptions on systematics, from $1 \ {\rm ab^{-1}}$ 
in $ep$ at the LHeC (`ep') and from the LHeC jointly
with CMS-S2 (`ep + pp')
\cite{LHeC:Higgs}. 
Empty bars represent cases
where no (or only a poor) measurement is possible. The Standard Model width is assumed. 
}
\label{fig:higgs}
\end{center}
\end{figure}

Higgs production in $ep$ 
scattering proceeds via the $WW \to H$ and $ZZ \to H$ subprocesses, easily 
distinguishable experimentally as the charged and neutral current processes, 
$ep \to \nu HX$ and $ep \to eHX$, respectively. 
In the final state, $b\bar{b}$ and $c\bar{c}$  
signatures are relatively easily identified in 
$ep$ scattering, without trigger rate issues and with controllable QCD backgrounds.
With a total Higgs production cross section of around 200 {\rm fb},
the LHeC when
running for a decade would produce hundreds of thousands 
of Higgs bosons. Detailed simulation studies have been performed, 
including realistic detector 
scenarios and background assessments, based on an integrated LHeC luminosity
of $1 \ {\rm ab^{-1}}$ \cite{LHeC:Higgs}. 
The results, both from the LHeC alone and in combination
with HL-LHC $pp$ data from CMS, are summarised in 
figure~\ref{fig:higgs}. 
Due to the dominance of the charged current process, the strongest sensitivity is to the
$WW$ Higgs coupling, which can be measured to 
significantly sub-1\% precision at the LHeC.
The projected LHeC sensitivities to the $ZZ$ and $b\bar{b}$ couplings 
are also at the 1\% level, whilst those to
$c\bar{c}$, $\tau\tau$ and $gg$ are sub-5\%. 
Longer-term targets with an FCC-eh programme  
include still stronger constraints on fermion and vector boson couplings as well as
probing for anomalies in the Higgs
self-coupling, with a $\sim \! 20\%$ measurement anticipated assuming Standard Model rates. 

\goodbreak
{\bf Future {\boldmath $ep$} facilities offer high sensitivities to several Higgs couplings, including some of those that are hardest to constrain at the HL-LHC. Concurrently operating {\boldmath $pp$} and {\boldmath $ep$} facilities at the LHC would therefore be a powerful medium-term combination. If the best possible precision is to be reached on Higgs couplings, each of the {\boldmath $pp$},  {\boldmath $ep$} and {\boldmath $e^+ e^-$} configurations have complementary roles to play.}

\null \vspace{-1.30cm}
\subsection{New Particles and Interactions}

Proposals for future lepton-hadron colliders with very high energy and luminosity 
are attracting significant theoretical attention for their
standalone potential sensitivity to BSM physics across a wide range of topics \cite{Azuelos:2018syu}.
The Standard Model Higgs programme described in section~\ref{sec:higgs} naturally extends to BSM 
scenarios. Encouraging studies are ongoing of LHeC and FCC-eh sensitivity to an extended Higgs sector
via Higgsino or charged Higgs production, as are studies of exotic Higgs decay
modes, including those involving dark matter particles. 

With a charged current single top production ($ep \rightarrow \nu t X$)
cross section of around $2 \ {\rm pb}$, the LHeC 
and FCC-eh are potential top 
factories that are particularly well suited to studying weak couplings
to the top quark and
probing for anomalies in
the $V_{tX}$ CKM matrix elements. The lack of top quarks in the proton PDFs
results in a very small neutral current single top cross section and corresponding strong sensitivity to flavor changing neutral currents. 

There is obvious sensitivity in lepton-hadron scattering to new `leptoquark' particles
with both lepton and quark quantum numbers (as appearing e.g. in R-parity violating SUSY)
that can be singly produced as resonances in the $s$-channel. 
If the leptoquark subsequently decays back to a lepton–quark system, 
the cross section depends quadratically on the coupling, and the
process is relatively easily identifiable through the signature of a resonance in the 
lepton plus jet invariant mass distribution.
Expectations for LHeC leptoquark sensitivity are 
comparable with the current LHC limits, reaching slightly beyond masses of $1 \ {\rm TeV}$ for
a coupling of electromagnetic strength. FCC-eh offers sensitivity up to $\sim \! 3 \ {\rm TeV}$
in the same scenario. 
Due to its very large  center-of-mass energy, VHEeP sensitivity approaches leptoquark
masses of $9 \ {\rm TeV}$ for sufficiently large couplings, significantly beyond the LHC expectations~\cite{Caldwell:2016cmw}.

\goodbreak
Following in the long tradition of resolving deeper structure of matter
through electron scattering that includes the discovery of 
quarks \cite{Bloom:1969kc,Breidenbach:1969kd}, 
{\bf
DIS at sufficiently large energy and luminosity is a natural environment in which to search for quark compositeness, reaching sensitivities below  $10^{-19}$~m, whilst charged current processes also offer sensitivity to heavy or sterile neutrino production. 
}

\section{Methodology and Readiness}
\label{methodology}

\subsection{Methodology} 
In view of the large and continuously growing realization of scientific promise, there are several proposals worldwide to build electron-hadron colliders. Various facilities,
with their key parameters listed in Table~\ref{tab:facilities},
are under consideration and will be further shortly described below, in order of increasing energy reach.
The facilities described here will require advances well beyond current particle accelerator technology.
This imposes both challenges and opportunities, as these advances are generally complementary to those
required for advancement of the energy frontier.
Examples include energy-recovery linear accelerators to achieve high-energy electron beams,
crab cavities for hadron beams,
and demonstration of achievability of high-energy physics-quality beams through wakefield acceleration.

\subsubsection{US-based Electron-Ion Collider (EIC) }

Two concepts for a future high-energy and high-luminosity polarized EIC have evolved in the United States. Both use existing infrastructure and facilities available to the US nuclear science community. Brookhaven National Laboratory (BNL) is proposing eRHIC through plans to add an electron storage ring to the existing RHIC Ion-Ion collider complex to enable electron-ion collisions. At Jefferson Lab (JLab), the Electron Ion Collider (JLEIC) design employs a new electron and ion collider ring complex, utilizing the 12-GeV upgraded CEBAF as electron injector. The high luminosity ($\sim\! 10^{34}$ cm$^{-2}$s$^{-1}$) requires the development of novel strong hadron cooling technologies,
which will push the limits of current technology and will therefore need significant R\&D. 
The EIC designs incorporate a wide range of beam energies and polarizations – both longitudinal and transverse - for protons, deuterons and helium-3 ions, including tensor polarization for deuteron beams – and ion species, enabling an unprecedented nucleon and nuclear femtography program

The EIC is foreseen to have at least two interaction regions (IRs). The physics-driven requirements on the EIC accelerator parameters, and extreme demands on kinematic coverage, makes integration into the IR of the main detector and dedicated detectors along the beamline to register all particles down to the smallest angles particularly challenging. The beamline detectors are fully integrated with the accelerator over a region of about 100 meters. 
To quickly separate both beams into their respective beam lines while providing the space and geometry required by the physics program, both designs incorporate a large crossing angle (of 20-50 mr). 
The detrimental effects of this crossing angle on the luminosity and beam dynamics are compensated by a crab-crossing scheme,
whose development has many synergies with the LHC high luminosity upgrade.
In the backward region an integrated magnet chicane allows electron detection down to the smallest angles to near-real photoproduction kinematics, inclusion of polarimetry systems, and an ancillary luminosity detector crucial for (1\%) precision measurement normalization. 

Driven by the demand for high precision on particle detection and identification of final state particles, modern particle detector and readout systems will be at the heart of the EIC. A multipurpose EIC detector needs full acceptance
and powerful hadron – lepton – photon separation and characterization.
It also requires excellent
pion, kaon and proton identification,
implying the use of
different particle identification technologies integrated over a wide rapidity range 
for particle momenta from a couple of 100 MeV to several tens of GeV.

\subsubsection{CERN-based Large Hadron-electron Collider (LHeC) }

The LHeC is an accelerator design to add intense 60 GeV electron beams based on an energy-recovery linear accelerator (ERL) to the LHC proton (and heavy ion) accelerator complex. Polarization of electron beams is possible in such an ERL-based scheme. The LHeC can provide electron-proton and electron-ion collisions running concurrently to the LHC operation, at very high energies, $\sqrt{s}$ = 1300 GeV (or $\sqrt{s}$ = 1800 GeV for the High-Energy LHC option), and high luminosity, $10^{34}$~cm$^{-2}$s$^{-1}$. 
This allows the LHeC to become a “giant microscope” of fast moving partons both to probe QCD and substructures at small momentum fractions, and to empower LHC into a new discovery era as a percent-level precision Higgs facility and for BSM model searches.

The principal advantage of the ERL-based linac-ring LHeC design is that the electron accelerator can be built, to considerable extent, 
independently of the LHC beam operations, unlike a ring-ring design. The ERL arrangement is located inside the LHC ring but outside the tunnel to minimize interference.  The wall-plug power of the default design has been constrained to 100 MW. In that configuration, two superconducting linacs, opposite to each other, accelerate the electrons in a 3-turn racetrack configuration to up to 60 GeV. Demonstration of the multi-pass and multi-GeV ERL technology in conditions of relatively large synchroton energy losses is the main challenge and is the subject of significant R\&D. The LHeC operation is transparent to the simultaneously running LHC experiments, owing to the low lepton bunch charge and resulting small beam-beam tune shift experienced by the proton beams. The LHeC would occupy one of the existing Interaction Points (IP); civil engineering options have been considered primarily for IP2. 

The LHeC detector has a classic collider detector structure, 
with an asymmetric design, taking into account the differences between the electron and proton/ion energies. In addition to the colliding electron and proton beams, the IR design has to accommodate the second non-interacting proton beam, to maintain concurrent LHeC and LHC operations. A combined magnet structure of central solenoid and forward dipoles has to be inserted between the electromagnetic and hadronic calorimeters. The central LHeC detector must be complemented by forward proton and neutron taggers while the backward region has to be instrumented with near-axis photon and electron detectors. These latter detectors are crucial to precisely determine the luminosity from Bethe-Heitler scattering. Key technology developments include large low-mass high-resolution tracking detectors, in connection with needs related to the LHC luminosity upgrade, but with largely reduced radiation and pile-up levels.

\goodbreak
\subsubsection{CERN-based Future Circular Collider with an electron accelerator (FCC-eh) }

The FCC is based on a 100 km circumference tunnel. The hadron-hadron collider design (FCC-hh) can be complemented by an electron accelerator. The FCC-eh design considers a similar ERL-based linac-ring design to the LHeC, but at point L on the FCC ring, not far from CERN itself.

The electron beam from the ERL would hit head on with one of the proton beams of the FCC while the other proton beam bypasses the IR. In a manner similar to the LHeC and LHC, the FCC-eh and FCC-hh are designed to operate concurrently. The FCC-eh would increase the energy reach for electron-proton collisions even more, to $\sqrt{s}$ = 3500 GeV, and thus extends the science reach of the LHeC “microscope” to even smaller momentum fractions. The envisioned luminosity given the higher proton energies would reach $1.5\times 10^{34} cm^{-2}s^{-1}$. 
In 25 years of concurrent ep operation at FCC, an integrated luminosity of 2-3 ab$^{-1}$ may be collected.
This energy and luminosity reach would allow an unprecedented laboratory for sub-percent precision Higgs measurements, proton structure to $x$ values as low as $10^{-7}$
and correspondingly enhanced discovery potential.

\subsubsection{CERN-based Very-High Energy electron-proton collider (VHEeP) }

The VHEeP design envisions  a 7-TeV LHC bunch as proton driver creating a wakefield which then accelerates electrons to 3 TeV. The 3 TeV electrons would then collide with the other LHC proton beam (a similar scheme, using the SPS as proton driver, is known as PEPIC). The foreseen energy reach of the VHEeP design is unprecedented, reaching $\sqrt{s}$ = 9000 GeV. The concept leads to modest luminosity, however, and is expected to be limited to $10^{29} cm^{-2}s^{-1}$. The VHEeP energy would constrain the high-energy behavior of hadronic cross sections and  reach the smallest momentum fractions, down to regions where new QCD dynamics must exist and connections with high-energy cosmic rays and gravity may be strongest.  

The VHEeP design makes strong use of current CERN infrastructure but would need a new tunnel to house the plasma accelerator. It relies on the concept that, similar as with laser pulses and electron bunches, a bunch of protons such as those of the LHC can create a wakefield plasma that can sustain very large electric fields capable of accelerating electron bunches. This is actively pursued by the AWAKE experiment at CERN, where it is investigated if bunches of electrons 
can be accelerated with high gradient and good emittance in 10 m of plasma. VHEeP operations would require separation of the drive proton beam and witness electron beam to avoid p-p collisions with the LHC beam: the beams will need to be separated transversely by typically one mm.

The VHEeP detector design would again require a central detector to reconstruct the hadronic final state combined with instrumentation close to the beam line to measure both electrons and hadrons. Spectrometers in the direction of both proton and electron beams will be required to measure the hadronic final state in certain regions and achieve close to hermeticity. Measurement of correlations of the struck quark remnants and the remaining hadronic remnants are likely not needed at the VHEep with its modest luminosity, as the emphasis will lie more with inclusive and diffractive final states, and total cross sections.

\subsection{Readiness and expected challenges}
Given the nature of these lepton-hadron colliders, all IR designs face the issues of strong focusing of beams at the collision point and fast separation of beams after the collision. The synchrotron radiation fan produced by the electrons in the various
arc dipoles and IR magnets has to be minimized and prevented from hitting beam pipes or materials in the vicinity of the detectors. Chromatic corrections related to the strong focusing have to be accomplished while maintaining acceptable dynamic aperture. Mitigation includes minimizing and softening the bend of the electron beams, inclusion of collimators, absorbers and masks at appropriate locations, and special design of the IR vacuum chamber and IR magnets.
Many lessons have been learned from previous lepton-lepton colliders and HERA, and many simulation tools exist, such that these considerations are sophisticatedly included in the various IR designs, and
not considered a serious challenge anymore. Beyond this challenge shared by all lepton-hadron colliders, we now list the readiness and main challenges facing the various lepton-hadron collider proposals – EIC, LHeC, FCC-eh and VHEep.

\subsubsection{EIC}

In recognition of the strong science case of the EIC, in 2015 the US Nuclear Science Advisory Committee (NSAC) advising the US Department of Energy (DOE) and the National Science Foundation, recommended an EIC in its Long-Range Plan as the highest priority for new construction \cite{EIC:LRP}. Subsequently, a National Academy of Sciences (NAS) panel was charged to review both the scientific opportunities enabled by an EIC and the benefits to other fields of science and society. The positive NAS report strongly articulates the merit of an EIC. The DOE Office of Nuclear Physics is already supporting increased efforts towards the most critical generic EIC-related accelerator R\&D. This concentrates on the development of strong hadron cooling technologies to underpin the luminosity needs, development of hadron beam crab cavities, R\&D on polarized ion sources and spin manipulation in the colliders, and further design of the IR magnets required. The EIC can be in operation in the late 2020's.

\subsubsection{LHeC}

The LHeC was principally designed in an extended conceptual 
report \cite{AbelleiraFernandez:2012cc}, published slightly before the Higgs boson 
discovery. A few years later CERN created the FCC studies, with an electron-proton collider as an option, and issued a mandate to develop the LHeC as an ERL to possibly be coupled with the LHC and its successors. The main technical challenge related to the LHeC is demonstration of the multi-pass and multi-GeV ERL technology in conditions of relatively large synchroton energy losses. This is a topic of worldwide current interest, with efforts underway at Cornell and Orsay. Work towards an LHeC electron accelerator prototype
(PERLE \cite{Angal-Kalinin:2017iup}), with two $80 \ {\rm MeV}$ ERL linacs based on 
$802 \ {\rm MHz}$ superconducting cavities already successfully built at JLab, is
underway at LAL, Orsay. Further ongoing R\&D is related to the high-current (polarized) electron source utilized in the linac-ring design, and the ability of the SRF cavities to operate with high average and peak beam currents, and the accompanying cryomodule design to contain the high beam power.
The latter questions have synergy with FCC needs. The LHeC can be in operation around 2030.

\subsubsection{FCC-eh}

The FCC-eh is the DIS option of the FCC, designed to be realized by adding the ERL, as developed and possibly
used for the LHeC, tangentially to the FCC hadron ring. Civil engineering studies prefer this to be realized
at point 'L' on the 100 km ring. The main accelerator challenge is coupled to the FCC-pp. If CERN decides for
the HE-LHC, instead of the FCC, to follow the LHeC, then the ERL developed for LHeC would directly serve for
ep and eA collisions.
The detector, be it for FCC-eh or HE-LHeC, is a scaled version of the LHeC detector. Its main challenge is
to achieve high-resolution coverage of especially the forward direction along the proton beam where both the
high-$Q^2$ backscattered electrons and the large-$x$ jets carry essentially the hadron beam energy.
Naturally, the time scale for the FCC-eh is linked to the FCC construction, and the FCC-eh can only be in operation in the mid-2040's.

\subsubsection{VHEeP}

The VHEeP design is strongly coupled to the concept of proton-driven wakefield acceleration. This is the topic of a proof-of-principle acceleration experiment at CERN called AWAKE. The first experiments focused on the self-modulation instability of the long proton bunch in the plasma, with promising results showing clear and reproducible self-modulation. Electron injection and acceleration up to 2 GeV was illustrated this year. The next stage is the study and technology demonstration of high gradient and good emittance electron acceleration, with a goal to be ready for final technology demonstration and application by 2024. The VHEeP can then be in operation in the early 2030s. A similar version using the SPS as proton driver (PEPIC) could potentially function as a demonstration project and be in operation in the mid to late 2020's.

\section{Summary}
\label{summary}

There has been tremendous progress over the last decade or so in the theoretical methodology to describe
deep inelastic scattering of leptons on hadrons. 
Coupled with new insights gained from experiments
worldwide, and strong synergy with recent advances in computing, 
this leads to a 
unique
opportunity for a decisive and transformational leap in our understanding of the strong interaction.
{\bf 
The facilities described here
are complementary and  are required to provide a complete characterization of the physics across the full kinematic reach.They
would lead to new, exciting, physics and a quantitative precision understanding.
Precision measurements in DIS would provide discovery potential and also sensitivity to the
direct production of new particles, in a cost-effective and timely manner.
}

In particular, the US-based EIC and the CERN-based LHeC are mature
and complementary projects, with extensive
efforts towards realization.
The EIC investigates the complex hadron-parton interface and would
be indispensable for performing detailed tomographic mapping of the hadrons' internal 
structure and for understanding the proton's mass and spin.
It would study the exact role quarks and gluons play in the structure and interactions of hadrons,
and 
nuclear binding. Using the nucleus as a probe or an amplifier, it would
constrain
the internal
wave function as relevant for the interpretation of CERN's heavy-ion experiments, and 
start studying the
role of multi-parton and collective effects in nuclei. The EIC science recently received strong endorsement
in a US National Academy of Sciences study.

The LHeC 
would establish DIS as part of the CERN accelerator complex
at the energy and luminosity frontier,
offering high precision measurements of the properties and couplings
of the Higgs boson and a considerable 
standalone programme of searches for
new physics.
It would also provide the cleanest high resolution microscope 
currently achievable to study the substructure and parton dynamics of protons,
empowering the CERN hadron collider programme through HL-LHC and beyond,
in particular maximizing its sensitivity to massive new particles. 
Furthermore, it
would transform our knowledge of the behavior of quarks and gluons in nuclei by hugely extending the
accessible kinematic range. 

For similar science reach, any Future Circular Collider should embed a lepton-hadron collider option,
FCC-eh, as an essential component included in the planning from the start. The discovery potential of
any new hadron-hadron collider will be amplified if beyond the direct production potential of colliders
at the energy frontier, this is coupled with a deeper understanding of the particles and forces that are
known. The energy reach of an FCC-eh would allow 
sub-percent precision Higgs measurements, enhanced standalone BSM sensitivity and a
map of longitudinal hadron 
structure over seven decades in $x$, correspondingly enhancing
discovery potential of FCC.

The 
FCC-eh and 
VHEeP energy would allow access to 
the smallest momentum fractions $x$, where
connections with high-energy cosmic rays and gravity may be strongest.
Measurements over such a tremendous energy range can reveal the 
new QCD dynamics at the heart of
high-energy hadronic cross sections. It is essential that the accelerator research and design related
to proton-driven wakefield acceleration, directly relevant for the VHEeP design, continue at rapid pace.
Progress is good: this year the CERN AWAKE experiment showed acceleration of electrons up to 2 GeV. With
continued focus towards technology demonstration of high gradient and good emittance electron acceleration,
CERN would be in an excellent position for a lepton-hadron collider demonstration project using this novel
technique.

\section*{Acknowledgments}
The authors of this document would like to acknowledge the many constructive comments from members of  the Deep Inelastic Scattering community that have helped shape and improve this document.

\newpage

\bibliographystyle{unsrt}
\bibliography{refs}

\end{document}